\documentclass[notitlepage,prd,twocolumn,showpacs,preprintnumbers,nofootinbib,superscriptaddress,floatfix,tightenlines]{revtex4-2}
\usepackage[utf8]{inputenc}
\usepackage{amsmath}    

\usepackage{mathrsfs}
\usepackage{amssymb}
\usepackage{graphicx}   
\usepackage{verbatim}   
\usepackage{color}      
\usepackage{xcolor}
\definecolor{lcolor}{rgb}{0.5,0,0}
\definecolor{citcolor}{rgb}{0,0,1}
\usepackage[breaklinks,colorlinks,urlcolor=blue,citecolor=citcolor,linkcolor=lcolor,linktoc=all]{hyperref}
\usepackage{bm}
\makeatletter
\g@addto@macro\bfseries{\boldmath}
\makeatother

\usepackage[capitalise]{cleveref}
\usepackage{slashed}
\usepackage{comment}

\newcommand{ \LambdaMS}{\bar{\Lambda}_{\mathrm{MS}}}

\newcommand{\KoKu}{SCC}
\newcommand{\ceft}{\chi}

\begin{document}
\author{Tyler Gorda}
\email{tyler.gorda@physik.tu-darmstadt.de}
\affiliation{Technische Universit\"{a}t Darmstadt, Department of Physics, 64289 Darmstadt, Germany}
\affiliation{ExtreMe Matter Institute EMMI, GSI Helmholtzzentrum f\"ur Schwerionenforschung GmbH, 64291 Darmstadt, Germany}
\author{Oleg Komoltsev}
\email{oleg.komoltsev@uis.no}
\author{Aleksi Kurkela}
\email{aleksi.kurkela@uis.no}
\affiliation{Faculty of Science and Technology, University of Stavanger, 4036 Stavanger, Norway}
\author{Aleksas Mazeliauskas}
\email{a.mazeliauskas@thphys.uni-heidelberg.de}
\affiliation{Institute for Theoretical Physics, University of Heidelberg, 69120 Heidelberg, Germany}
\title{\texorpdfstring{Bayesian uncertainty quantification of perturbative QCD input \\to the neutron-star equation of state}{Bayesian uncertainty quantification of perturbative QCD input to the neutron-star equation of state}}
\begin{abstract}
    The equation of state of neutron-star cores can be constrained by requiring a consistent connection to the perturbative Quantum Chromodynamics (QCD) calculations at high densities. The constraining power of the QCD input depends on uncertainties from missing higher-order terms, the choice of the unphysical renormalization scale, and the reference density where QCD calculations are performed. Within a Bayesian approach, we discuss the convergence of the perturbative QCD series, quantify its uncertainties at high densities, and present a framework to systematically propagate the uncertainties down to neutron-star densities. We find that the effect of the QCD input on the neutron-star inference is insensitive to the various unphysical choices made in the uncertainty estimation.  
\end{abstract}



%

\maketitle
\section{Introduction}

The determination of the equation of state (EoS) of neutron-star (NS) cores is one of the grand questions of nuclear astrophysics~\cite{Lovato:2022vgq,Sorensen:2023zkk}. The EoS determines many of the macroscopic properties of neutron stars and its features may give a unique inroad into determining the phase structure of Quantum Chromodynamics (QCD) at large baryon number densities \cite{Annala:2019puf, Fujimoto:2022ohj, Kojo:2021ugu,Tan:2021nat}.

In the past years there has been an extremely rapid evolution in NS observations, e.g.~\cite{Demorest:2010bx, Antoniadis:2013pzd,Nattila:2017wtj, LIGOScientific:2017vwq, LIGOScientific:2018cki,Linares:2018ppq, NANOGrav:2019jur, Fonseca:2021wxt,  Miller:2019cac,Riley:2019yda,Miller:2021qha, Riley:2021pdl, Romani:2022jhd}, combined with maturing theoretical and statistical techniques, e.g.~\cite{Landry:2018prl,Essick:2019ldf,Miller:2019nzo,Fujimoto:2019hxv,Raaijmakers:2019dks, Miller:2019nzo, Essick:2020ghc,Essick:2020flb,Al-Mamun:2020vzu,Dietrich:2020efo,Huth:2021bsp,Raaijmakers:2021uju, Essick:2021kjb,  Legred:2022pyp, Gorda:2022jvk,Han:2022rug, Jiang:2022tps}, to constrain and infer the EoS using a variety of observational and theoretical inputs. Among the theoretical inputs are the \emph{ab-initio} calculations determining the EoS directly from the Lagrangian of QCD using perturbation theory
\cite{Freedman:1976ub, Fraga:2004gz, Kurkela:2009gj, Kurkela:2016was, Gorda:2018gpy, Gorda:2021gha,Gorda:2021znl, Gorda:2021kme}. These calculations rely on the asymptotic freedom of QCD dictating that at high densities the EoS can be expanded in powers of the strong coupling constant $\alpha_s$. 
   
At sufficiently high densities, well above the density range reached in stable NSs, perturbation theory gives a good approximation of the true EoS. It has furthermore been recently shown that these calculations---combined with the requirement that the EoS be mechanically stable, causal, and thermodynamically consistent (\KoKu) at all densities---give robust constraints to the EoS down to a few saturation densities $n \sim 2.3 n_s$ \cite{Komoltsev:2021jzg}, with $n_s \approx 0.16\,\mathrm{fm}^{-3}$. The interaction between the astrophysical  and the QCD constraints has also been studied, showing that the QCD input leads to a softening of the EoS at the highest densities reached inside the cores of stable NSs \cite{Kurkela:2014vha,Annala:2019puf,Ecker:2022xxj,Fujimoto:2022ohj,Marczenko:2022cwq,Altiparmak:2022bke,Gorda:2022jvk,Somasundaram:2021clp} (cf. \cite{Somasundaram:2022ztm}). This feature has been interpreted as a sign of loss of hadronic structure, and a phase change to quark matter \cite{Annala:2019puf, Fujimoto:2022ohj, Kojo:2021ugu, Han:2022rug}. 
The importance of the theoretical inputs in the EoS inference necessitates reliable and statistically interpretable  uncertainty estimation of the calculations.
In the low-density nuclear regime, theoretical uncertainty estimation including statistical uncertainty quantification has been an increasingly studied problem in recent years \cite{Furnstahl:2014xsa,Wesolowski:2015fqa,Drischler:2017wtt,Lim:2018bkq,Wesolowski:2018lzj,Melendez:2019izc,Drischler:2020hwi,Drischler:2020yad,Keller:2020qhx,Keller:2022crb,Elhatisari:2022qfr,Lovato:2022vgq}. Such quantified uncertainties are now routinely incorporated when inferring the NS-matter EoS.  With this paper, we improve the treatment of corresponding uncertainties from the perturbative-QCD (pQCD) input used in the inference to produce statistically interpretable error estimates of the pQCD results and their impact at low densities.

The accuracy of the pQCD calculation is limited by the ignorance of the terms in the perturbative series beyond the last computed order, i.e., the missing-higher-order (MHO) terms.  The current standard has been 
to estimate the MHO uncertainty of the result through its variation with respect to an unphysical renormalization scale $\bar{\Lambda}$. Explicitly, within the $\LambdaMS$ renormalization scheme, $\bar{\Lambda}$ is varied by a factor of two around a central scale of $2\mu/3$, with $\mu$ the baryon chemical potential, to produce an uncertainty band for the result \cite{Fraga:2004gz,Kurkela:2009gj,Kurkela:2016was,Gorda:2018gpy,Gorda:2021gha}. 
While this \emph{ad-hoc} procedure is rooted in historical practice and experience in perturbative calculations, it lacks a well-defined statistical interpretation.

Recently, a Bayesian approach to estimate perturbative uncertainties has received attention in the high-energy community producing predictions for LHC physics \cite{Cacciari:2011ze,Forte:2013mda,Bagnaschi:2014wea,Bonvini:2020xeo,Duhr:2021mfd}. 
These studies have differentiated between scale-variation uncertainties (i.e.,~those arising from setting the unphysical renormalization scale) and MHO corrections (i.e.,~those arising from the truncation of the perturbative series). 
For the MHO errors, various machine-learning-based models have been suggested to synthesize information from all the computed partial sums of the perturbative series (instead of solely from the highest order). Uncertainties due to scale-variation can also be folded in using different marginalization procedures.

In this work, we explore these different possibilities for quantifying the uncertainties of the high-density pQCD EoS at different chemical potentials $\mu$ using the \texttt{MiHO} code~\cite{Duhr:2021mfd,mihogit}.
Additionally, we further develop the statistical framework to propagate the pQCD results to NS densities \cite{Komoltsev:2021jzg, Gorda:2022jvk}. In particular, we discuss the marginalization of the chemical potential where the pQCD result is used. In this way we can combine the information of how constraining the pQCD results are with how convergent the pQCD series is at different chemical potentials. This extends the previous works that have so far considered only a fixed chemical potential for applying the pQCD results. 
The organization of the paper is outlined in the following Section. 

%

\section{Overview of the setup}
\label{sec:setup}

 In this section we present the overview of our setup of first estimating the perturbative uncertainties of the EoS at high densities using the Bayesian machine-learning-based framework \texttt{MiHO}~\cite{Duhr:2021mfd, mihogit} and then applying robust equation-of-state constraints~\cite{Komoltsev:2021jzg, Gorda:2022jvk} to translate the high-density information to NS densities.The different elements of the framework are introduced here while the details are discussed in the following sections.
 
 Our goal is to determine the posterior probability 
 \begin{equation}
 P(\epsilon_L, p_L | n_L, \bm{p}^{(k)},\bm{n}^{(k)} )
 \end{equation} 
 of the (reduced) EoS $p_L(\epsilon_L)$ at NS densities $n_L$, given the first $k+1$ terms of the perturbative series for the pressure\footnote{We refer here to the first $k+1$ terms being summed and not the first $k+1$ partial sums.}  $\bm{p}^{(k)} \equiv (p^{(0)}, \cdots,  p^{(k)})$ and number density $\bm{n}^{(k)} \equiv (n^{(0)},\ldots,n^{(k)})$ at high densities where the perturbative description of QCD is reliable. Here, $\epsilon_L$, $p_L$, and $n_L$ are the energy density, pressure, and baryon number density to which we wish to propagate the perturbative input; that is, at and around NS densities. The subscript $L$ refers to low densities as opposed to the high densities, $H$, where the pQCD results converge and are directly valid.

The first step is to convert the information about the perturbative series of the pressure $p^{(k)}$ and particle-number density  $n^{(k)}$ to a statistically interpretable probability distribution of pressure $p_H$ and density $n_H$ at a given large baryon number chemical potential $\mu_H$. To this end we use statistical machine-learning models that bound the higher-order terms $\{p^{(k + 1)}, p^{(k + 2)}, \ldots\}$ given the lowest $k+1$ terms. We denote the resulting joint probability distribution quantifying the remaining MHO uncertainty as
\begin{equation}
    P_{\rm MHO}( p_H, n_H | \bm{p}^{(k)}(\mu_H , X), \bm{n}^{(k)}(\mu_H , X)),\label{eq:jointPnp}
\end{equation}
where $X = 3\bar \Lambda/(2\mu)$ is related to the unphysical renormalization scale $\bar \Lambda$ upon which the perturbative coefficients depend. We use the \texttt{MiHO} computer code to estimate these probabilities~\cite{mihogit}. 
The current state-of-the-art perturbative coefficients $p^{(k)}$ are reported in \cref{sec:pQCD} and the details of the probability distribution of \cref{eq:jointPnp} is discussed in \cref{sec:MiHO_A}. 

In order to remove the dependence on the unphysical renormalization scale $X$, one is forced to integrate over a sufficiently wide range of scales
\begin{align}
\label{eq:sm/sa}
  P( p_H, & n_H | \bm{p}^{(k)}(\mu_H), \bm{n}^{(k)}(\mu_H)) \\ & = \int\! dX\,  P_{\rm MHO}( p_H, n_H | \bm{p}^{(k)}(\mu_H , X), \bm{n}^{(k)}(\mu_H , X)) \nonumber \\
  & \quad\quad\quad \times P_{\text{sa/sm}}( X |\bm{p}^{(k)}(\mu_H), \bm{n}^{(k)}(\mu_H)),\nonumber
\end{align}
where $P_{\text{sa/sm}}( X |\bm{p}^{(k)}(\mu_H), \bm{n}^{(k)}(\mu_H))$ is an integration weight to be determined by a specific prescription. In this context we will discuss the scale-marginalization (sm) and scale-averaging (sa) prescriptions introduced in \cite{Bonvini:2020xeo} and \cite{Duhr:2021mfd}, respectively; for details, see \cref{sec:MiHO_B}. 
This procedure is applied to the known non-perturbative results at high temperatures and zero chemical potential in \cref{sec:high_T}. 

Given the EoS at high chemical potentials $\mu_H$ --- that is, the triplet of values $p_H, n_H$, and $\mu_H$ --- we can determine the region of allowed values $\epsilon_L, p_L$ at some lower density $n_L$ using the robust equation-of-state constraints introduced in \cite{Komoltsev:2021jzg}. These robust constraints are based on considering the most extreme extrapolations of the pQCD EoS that are allowed by mechanical stability, causality, and thermodynamic consistency to lower densities relevant to neutron stars. They can be expressed as a conditional probability\footnote{We note that we use different thermodynamic variables at high and low densities. Namely, at low densities we characterized the EoS based on density $n_L$ whereas we use the chemical potential $\mu_H$ at high densities. The reason for this choice is that the chemical potential is the natural variable in the pQCD calculation whereas the density is more directly relevant for the physics of NSs.  

}
\begin{align}
\label{eq:koku}
    P_{\rm \KoKu}(\epsilon_L, p_L |  n_L, \mu_H , p_H, n_H ),
\end{align}
which takes a value 1 if the low density EoS at ($n_L$, $\epsilon_L$, $p_L$) can be connected to the high density one at ($\mu_H$, $n_H$, $p_H$) with \emph{any} stable, causal, and consistent (SCC) extension and 0 otherwise. For details see \cref{sec:KoKu}.
Combined with the joint probability distribution for the high-density EoS in \cref{eq:jointPnp}, we can write down the distribution for $\epsilon_L$ and $p_L$ at some given density $n_L$ and a matching scale $\mu_H$
\begin{align}
     P(\epsilon_L,& p_L |  { \bm p}^{(k)}, { \bm n}^{(k)}, n_L, \mu_H, X)  \nonumber \\
 =&\int dp_H dn_H  P_{\rm \KoKu}(\epsilon_L, p_L | n_L,\mu_H,p_H, n_H)   \nonumber \\
  &\times   P_{\rm MHO}(p_H,  n_H |\bm{p}^{(k)}(\mu_H , X), \bm{n}^{(k)}(\mu_H , X)).
\end{align}
Finally, since the matching scale $\mu_H$ is in principle arbitrary (as long as perturbative calculations remain reliable), we may marginalize over possible values of it similarly to what is done for the renormalization scale; 
the details are given in \cref{sec:KoKu_B}.
The complete formula is then given by
\begin{align}
\label{eq:master_formula_full}
P(&\epsilon_L, p_L | n_L, \bm{p}^{(k)},\bm{n}^{(k)})\nonumber \\
& = \int d\mu_H dp_H dn_H dX \nonumber   \\ 
& \quad  \times  P_{\rm \KoKu}(\epsilon_L, p_L |  n_L, \mu_H , p_H, n_H ) \ \nonumber  \\
& \quad \times P_{\rm sa/sm}(\mu_H, X |\bm{p}^{(k)}, \bm{n}^{(k)}) \nonumber \\
&  \quad \times   P_{\rm MHO}(p_H, n_H| \bm{p}^{(k)}(\mu_H , X),\bm{n}^{(k)}(\mu_H , X)).\end{align}

While optimally we would use the above expression to estimate the final posterior probability distribution, for practical reasons we have to make certain well-justified simplifications to it. 
Currently the \texttt{MiHO} computer code produces posterior distributions of one variable and is unable to compute joint probability distributions of two variables as required by \cref{eq:jointPnp}. To estimate the joint probability distribution, we conservatively assume independent  distributions for $p_H$ and $n_H$ such that  
\begin{align}
\label{eq:uncorrelated}
&P_{\rm MHO}( p_H, n_H | \bm{p}^{(k)}(\mu_H , X), \bm{n}^{(k)}(\mu_H , X))  \nonumber \\&\,\,\approx P_{\rm MHO}(p_H | \bm{p}^{(k)}(\mu_H , X))P_{\rm MHO}(n_H |  \bm{n}^{(k)}(\mu_H , X))\nonumber \\
&\,\,\approx P_{\rm MHO}(p_H | \bm{p}^{(k)}(\mu_H , X))\delta(n^{(k)}(\mu_H,X)-n_H),
\end{align}
where we have also used the fact that the series for $n$ converges faster than the series for $p$ in pQCD.
Similarly, we estimate the integral weights arising from the scale marginalization procedure using only information about the perturbative series in pressure
\begin{align}
P_{\rm sa/sm}(\mu_H, X |\bm{p}^{(k)}, \bm{n}^{(k)}) \approx P_{\rm sa/sm}(\mu_H, X |\bm{p}^{(k)} ).\label{eq:finalweight}
\end{align}
Both of these approximations are well justified given that the perturbative series for the number density $\bm{n}^{(k)}$ converges faster than the corresponding series for the pressure $\bm{p}^{(k)}$. We leave improvements to the \texttt{MiHO} computer code and associated models discussed below in \cref{sec:MiHO_A} to future work.

With these approximations the final result can be written as
\begin{align}
\label{eq:master_formula}
P(\epsilon_L, p_L &| n_L, \bm{p}^{(k)})\nonumber \\
& = \int d\mu_H dp_H dn_H dX \nonumber   \\  
& \quad  \times  P_{\rm \KoKu}(\epsilon_L, p_L |  n_L, \mu_H , p_H, n_H ) \ \nonumber  \\
& \quad \times P_{\rm sa/sm}(\mu_H, X |\bm{p}^{(k)}) \nonumber \\
&  \quad \times   P_{\rm MHO}(p_H| \bm{p}^{(k)}(\mu_H , X)) \nonumber \\
& \quad  \times \delta(n^{(k)}(\mu_H,X)-n_H)  
\end{align}

In the next three sections, we discuss the three types of input used in the master formula \cref{eq:master_formula}.

\section{pQCD at high densities}
\label{sec:pQCD}
In this section we discuss the current state-of-the-art pQCD results for the EoS at zero temperature $T$ in $\beta$-equilibrium. We  limit the discussion to 3 flavors of massless quarks, for which the $\beta$-equilibrium condition trivializes and the individual quark chemical potentials are given by $\mu_u = \mu_d = \mu_s = \mu/3 \equiv \mu_q$, with $\mu$ being the baryon-number chemical potential. The approximation of zero quark masses is justified as we use the pQCD result only at very large chemical potentials, much larger than the strange quark mass $\mu_q \gg m_s$, such that the corrections from the strange quark masses are small \cite{Kurkela:2009gj,Fraga:2004gz,Gorda:2021gha}. At the same time, the chemical potential is below the charm threshold  justifying restricting to 3 flavors. 

\begin{figure*}
    \centering
    \includegraphics[width=2\columnwidth]{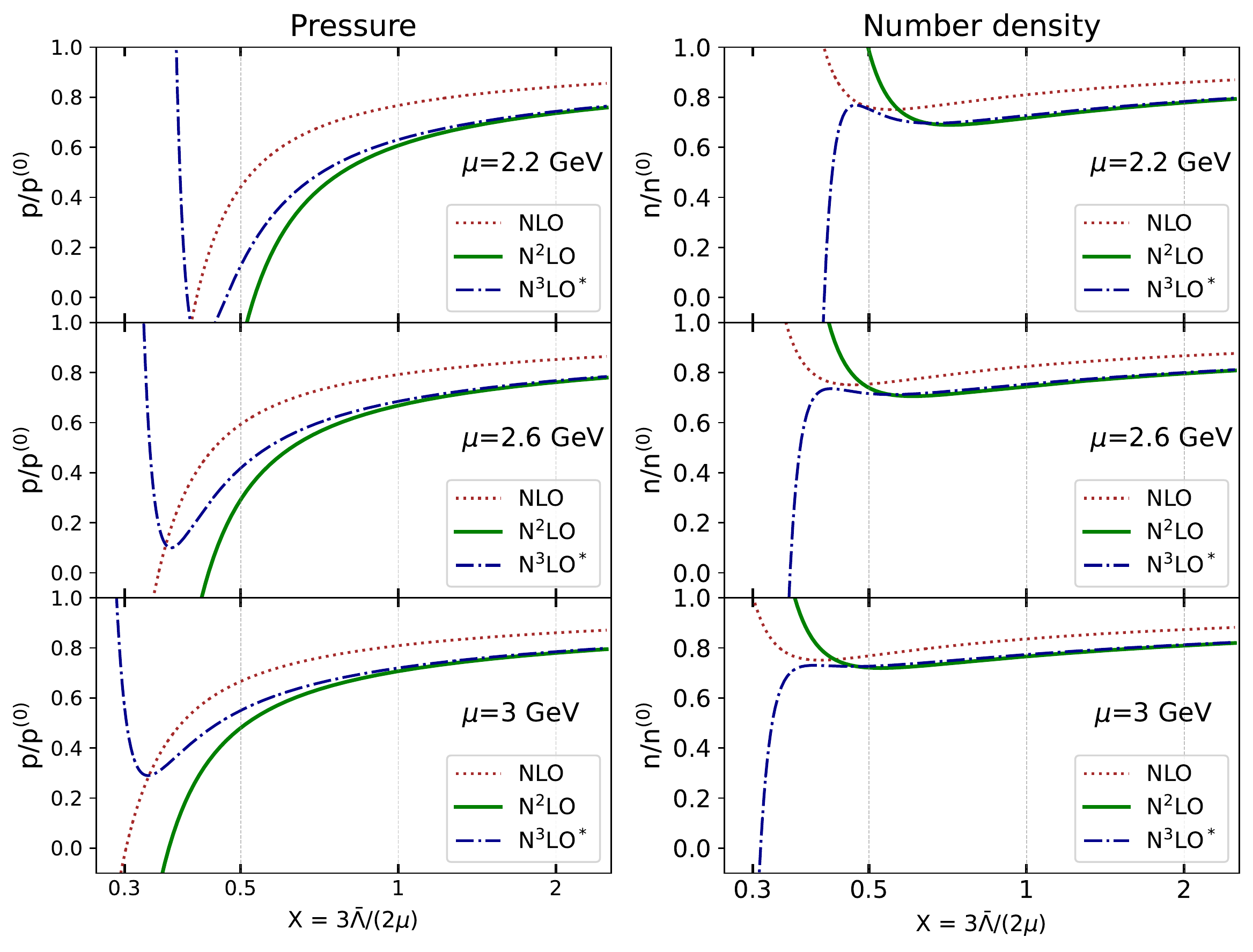}   
    \caption{Order-by-order perturbative-QCD results for (left) the normalized pressure  and (right) normalized density as a function of renormalization scale  [see \cref{eq:pseries}] at different chemical potentials $\mu = \{2.2, 2.6, 3.0\}$GeV, corresponding roughly to densities of $n \approx \{23, 40, 63\}n_s$, respectively.}
    \label{fig:pQCD_scale-dep}
\end{figure*}

In the following we consider only the results for the pressure because the pressure and density are linked to each other through the thermodynamic identity 
\begin{align}
     n(\mu) \equiv \left.\frac{\partial p(\mu) }{\partial \mu} \right|_{T}.
\end{align}

The perturbative expansion for the pressure reads
\begin{align}
    p(\mu) \simeq p_n(\mu, \bar \Lambda ) \equiv \sum_{k=0}^n  p^{(k)}(\mu,  \bar \Lambda)\label{eq:pseries}
\end{align}
where the different coefficients depend on $\bar \Lambda$ explicitly as well as through the strong coupling constant $\alpha_s(\bar \Lambda)$. 
The leading-order (LO) and next-to-leading-order (NLO) results are easily extracted as the $T\to 0$ limit of results in \cite{Kurkela:2016was}
\begin{align}
    p^{(0)} & = \frac{C_A N_f }{12 \pi^2 }\mu_q^4
    \overset{\rm QCD}= \frac{3 \mu_q^4}{4\pi^2 }\ , \\
    p^{(1)} & = - \alpha_s(\bar \Lambda)\frac{3 d_A }{4 C_A \pi} p^{(0)} \overset{\rm QCD}= -\frac{2}{\pi} \alpha_s(\bar \Lambda) p^{(0)},
\end{align}
where $C_A = N_c = 3, d_A = N_c^2 - 1 = 8$ are the usual color factors and $N_f=3$ is the number of active fermion flavors.

The next-to-next-to-leading-order (N$^2$LO) result consists of two types of terms,
\begin{align}
    p^{(2)} &= p_{\rm hard}^{(2)} + p_{\rm soft}^{(2)} \label{eq:N2LOp}.
\end{align}
The first one is the hard contribution\footnote{We note that in Eq.~(B7) of \cite{Kurkela:2016was}, there is a factor $3/2$ missing from in the second term. This factor is present in Eq.~(B6).  },
\begin{align}
   p_{\rm hard}^{(2)} = &-\alpha_s^2(\bar \Lambda) \frac{d_A}{96 C_A \pi ^2} \Big[284 C_A-153 C_F 
   \nonumber \\ & +2 N_f \left( 11-36 \gamma_E +48\log (2) \right) 
   \nonumber \\ & +12 (11 C_A-2 N_f) \log \left(\frac{\bar \Lambda}{2 \mu_q}\right) 
   \nonumber \\& -72 N_f \log \left(\frac{\mu_q}{2 \pi  T}\right)\Big] p^{(0)},
\end{align}
that arises from the 3-loop bubble diagrams. In these diagrams, the momenta are of order $\mu_q$. The second term in \cref{eq:N2LOp} is due to the soft divergence of QCD that is regulated by the in-medium screening of color charges, which leads to an enlarged set of diagrams with soft momenta $\sim \alpha_s^{1/2}\mu_q$. In this case, one must perform an all-loop-orders resummation of a particular class of ``ring diagrams'' to correctly capture all of the physics from this soft momentum scale at this order in the coupling; for details, see \cite{Gorda:2021kme}.  The resummation is most conveniently performed using the hard-loop effective theory~\cite{Ghiglieri:2020dpq} leading to the result \cite{Kurkela:2016was}
\begin{align}
  p_{\rm soft}^{(2)} = \frac{d_A m_E^4}{128 \pi ^2} \left[\log \left(\frac{T^2}{m_E^2}\right)+ 6.6719\right],
\end{align}
where $m_E^2 = \frac{2 \alpha_s(\bar\Lambda)}{\pi} N_f \mu_q^2$ is the in-medium screening mass scale. 

Separately the two contributions in \cref{eq:N2LOp} are divergent at zero temperature, but the $T$ dependence vanishes in the sum of the two contributions. Only the hard contribution has a physical UV divergence that is regulated by renormalization. Therefore, only the hard contribution explicitly depends on the unphysical renormalization scale $\bar \Lambda$. The dependence on $\bar \Lambda$ appears naturally in the logarithm together with the scale $2\mu_q$. This factor of 2 can be traced to the coefficient of $\epsilon$ in performing Feynman integrals with integer powers of propagators in $D= 4-2\epsilon$ spacetime dimensions in the $\LambdaMS$-scheme. This is the motivation for choosing the central renormalization scale $\bar \Lambda = 2\mu_q = \frac{2}{3}\mu$. The dimensionless quantity parameterizing the variation around the central scale is denoted by $X$, such that $\bar \Lambda =  2 \mu_q X$, and the conventional scale variation by a factor of two is given by varying $X \in [1/2,2]$.\footnote{Note that in other works (cf.~\cite{Kurkela:2009gj,Kurkela:2014vha,Gorda:2021znl}), a  different convention for $X$ is used, namely $\bar\Lambda = \mu_q X'$, with $X' \in [1,4]$.}

At N$^3$LO only the soft contribution to the pressure is known (this time at exactly zero $T$) \cite{Gorda:2021kme, Gorda:2021znl}, 
\begin{align}
    p_{\rm soft}^{(3)} =0.26587 \frac{\alpha_s(\bar\Lambda) C_A d_A m_E^4}{(8 \pi)^2}.
\end{align}
We will not use this result unless explicitly stated, and when doing so we will denote this order with an asterisk ---  N$^3$LO$^*$ --- to remind the reader that not all contributions at this order are included. 

In our results, we use the above expressions for the EoS, including the two-loop running of the strong coupling $\alpha_s(\bar{\Lambda})$  in all results. We set the non-perturbative scale $\Lambda_\mathrm{QCD} = 378$~MeV, which fixes $\alpha_s(2\text{ GeV}) = 0.2994$.
The renormalization-scale dependence of different perturbative orders of pressure and number density is depicted in \cref{fig:pQCD_scale-dep} at different values of the baryon chemical potential. The fiducial range of scale variation is shown as vertical lines. We note that for all but the N$^3$LO$^*$ results, there exists a smallest value of $X$ for which the pressure remains positive. As $\mu$ is decreased, this smallest value of $X$ increases, eventually approaching or moving into the fiducial range. We note that this diverging behavior at small $X$ is the origin of sizeable scale variation errors in the conventional practice. The number density has a much smaller dependence on $X$ and remains positive in the fiducial range even at $\mu=2.2\,\text{GeV}$.

\section{Estimating MHO and renormalization-scale uncertainties}
\label{sec:MiHO}

In this section, we discuss our handling of the uncertainties of the high-density QCD EoS. On the one hand, the truncated perturbative expansion \cref{eq:pseries} differs from the actual value by MHO terms. On the other hand, the truncated series depends on the choice of the renormalization scale, which is controlled by the parameter $X$.
We use the recently introduced \texttt{MiHO} framework to estimate both uncertainties using Bayesian machine-learning techniques~\cite{Duhr:2021mfd}. 
The main idea of the framework is to assume that the perturbative coefficients can be taken as independent draws from distributions arising from a statistical model of convergent series.
Performing Bayesian inference on the available perturbative orders allows one to constrain the model parameters and construct the probability distribution for the next term in the series. The spread of the posterior distribution can then be used to quantify the uncertainty of the MHO terms. The scale uncertainty is incorporated by combining the probability distributions at different scales using a particular scale prescription, which will be specified below.

\subsection{Posterior distribution at fixed scale}
\label{sec:MiHO_A}

As explained in \cref{sec:setup} [see \cref{eq:uncorrelated}], we desire the probability distribution of the pressure and number density given the first $k+1$ terms in the perturbative expansion $\bm{p}^{(k)}=(p^{(0)}(\mu,X),\ldots p^{(k)}(\mu,X))$ for fixed $\mu$ and $X$ %
\begin{align}
    P_{\rm MHO}(p | \bm{p}^{(k)}(\mu,X)).
    \label{eq:miho_prob}
\end{align}
For a convergent series the probability for the sum can be approximated by the posterior probability for the next partial sum in the series, which we will infer using two models: the geometric and $abc$ models.

In practice we work with perturbative corrections normalized to the LO term. That is, we define a sequence of coefficients 
\begin{equation}
 \delta_k(\mu,X)= \frac{p^{(k)}(\mu,X)}{p^{(0)}(\mu,X)}
\end{equation}
and $\bm{\delta}_k \equiv (\delta_0, \ldots, \delta_{k-1}, \delta_k)$ with $\delta_0=1$ by definition.
The Bayesian model consists of a parameterized prior distribution for each $\delta_k$. 

The \textbf{geometric model} assumes a flat distribution for 
\begin{equation}
\frac{|\delta_k|}{a^k}\leq c
\end{equation}
with hidden model parameters $a$ and $c$~\cite{Bonvini:2020xeo}. The normalized prior distribution for the $\delta_k$ is
\begin{equation}
    P_\text{geo}(\delta_k|a,c) \equiv \frac{1}{2a^kc}\theta\left(c-\frac{|\delta_k|}{a^k}\right).\label{eq:geo}
\end{equation}
That is, the sequence of perturbative coefficients is considered to be random variables, drawn from a uniform distribution $\delta_k \sim \mathcal{U}_{[ - c a^k , c a^k ]}$ whose width decreases geometrically with increasing order for $0<a<1$.

The \textbf{$abc$ model} introduced in~\cite{Duhr:2021mfd} was proposed to allow for an asymmetric distribution
\begin{equation}
b-c \leq \frac{\delta_k}{a^k} \leq b+c
\end{equation}
 with
\begin{equation}
    P_{abc}(\delta_k|a,b,c) \equiv \frac{1}{2|a|^kc}\theta\left(c-\left|\frac{\delta_k}{a^k}-b\right|\right)\label{eq:abc}.
\end{equation}
Note that in the $abc$ model, the parameter $a$ can take negative values ( $-1<a<1$), and therefore it can differentiate between alternating and non-alternating series.

The model is then trained using the sequence of known terms, leading to posterior distributions for the model parameters $a$, ($b$), and $c$ using Bayes's theorem
\begin{align}
    P(a,c| \bm{\delta}_k) = 
    \frac{P( \bm{\delta}_k | a, c) P_0(a)P_0(c)}
    {P(\bm{\delta}_k)},
\end{align}
where $P( \bm{\delta}_k | a, c)$ is the product of \cref{eq:geo} [or \cref{eq:abc}] for each term in $\bm{\delta}_k$ and $P_0(a)$ and $P_0(c)$ are judiciously chosen priors.

For the geometric model $P_0(a)$ and $P_0(c)$ are chosen to satisfy $0<a<1$ and $c\geq 1$:
\begin{align}
    P_0(a)&\equiv(1+\omega)(1-a)^\omega \theta(a)\theta(1-a),\\
    P_0(c)&\equiv\frac{\epsilon}{c^{1+\epsilon}}\theta(c-1).
\end{align}
Here $\epsilon=0.1$ and $\omega=1$ are default constants defining the prior distributions. 

For the $abc$ model the priors are chosen as
\begin{align}
    P_0(a)&\equiv\frac{1}{2}(1+\omega)(1-|a|)^\omega \theta(1-|a|),\\
    P_0(b,c)&\equiv\frac{\epsilon \eta^\epsilon}{2\xi c^{2+\epsilon}}\theta(c - \eta)\theta(\xi c-|b|).
\end{align}
The default constants for the $abc$ model are $(\epsilon, \omega, \xi, \eta)=(0.1,1,2,0.1)$.\footnote{The sensitivity to the choices of these constants was studied in~\cite{Duhr:2021mfd}. Generically, the sensitivity to priors reduces with increasing perturbative order.}

The marginalized likelihood (or evidence) for the geometric model is given by
\begin{align}
    P(\bm{\delta}_k) \equiv \int da dc P(\bm{\delta}_k| a, c) P_0(a)P_0(c),\label{eq:evidence}
\end{align}
and similarly for the $abc$ model. The marginalized distribution measures the (relative) confidence in the model's capabilities to reproduce the known terms in the series.

We use the trained model to find the posterior distribution of the next order in the series
\begin{align}
P(\delta_{k + 1}| \bm{\delta}_k) =
\int da dc P(\delta_{k+1}| a, c ) P(a, c | \bm{\delta}_k).
\end{align}
Then our desired posterior probability distribution for the pressure is approximated by~\cite{Duhr:2021mfd}
\begin{align}
    \label{eq:PMHO}
    P_{\rm MHO}(p | \bm{p}^{(k)}(\mu,X)) \approx \frac{1}{p^{(0)}} P\left(\delta_{k+1}=\frac{p - \sum_{i=0}^k p^{(i)}}{p^{(0)}}| \bm{\delta}_k\right).
\end{align}
\begin{figure*}
    \centering
    \includegraphics[width = 0.9 \textwidth]{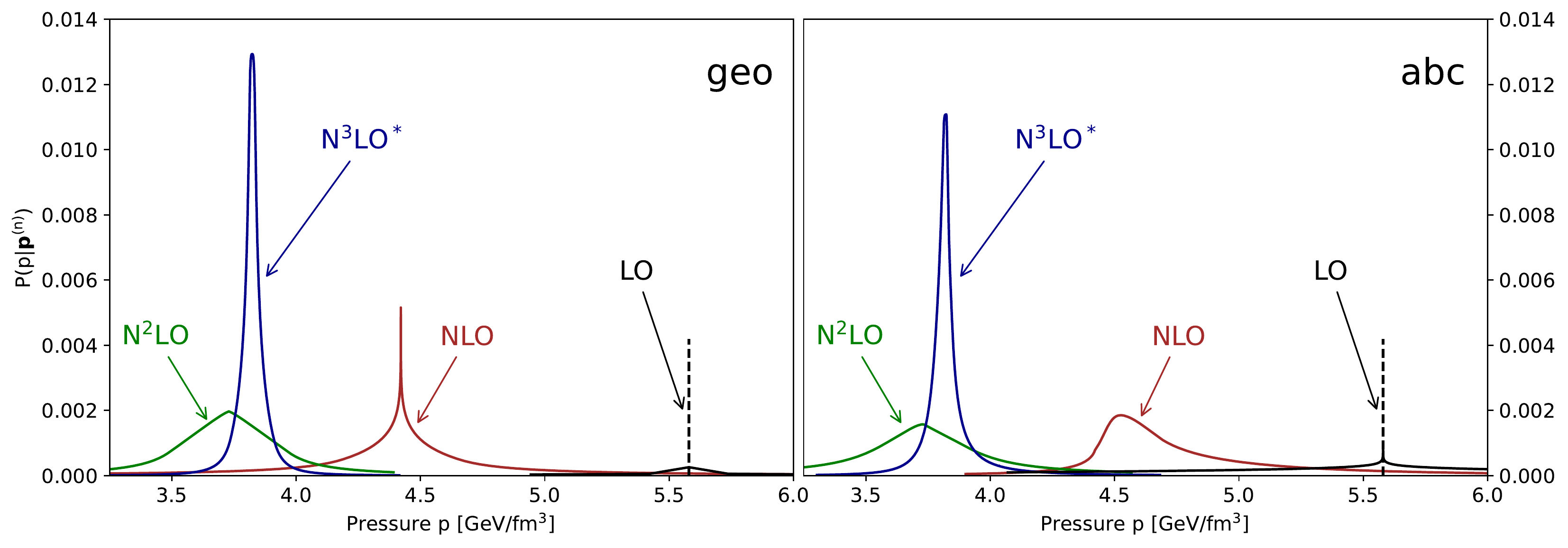}
    \includegraphics[width = 0.9 \textwidth]{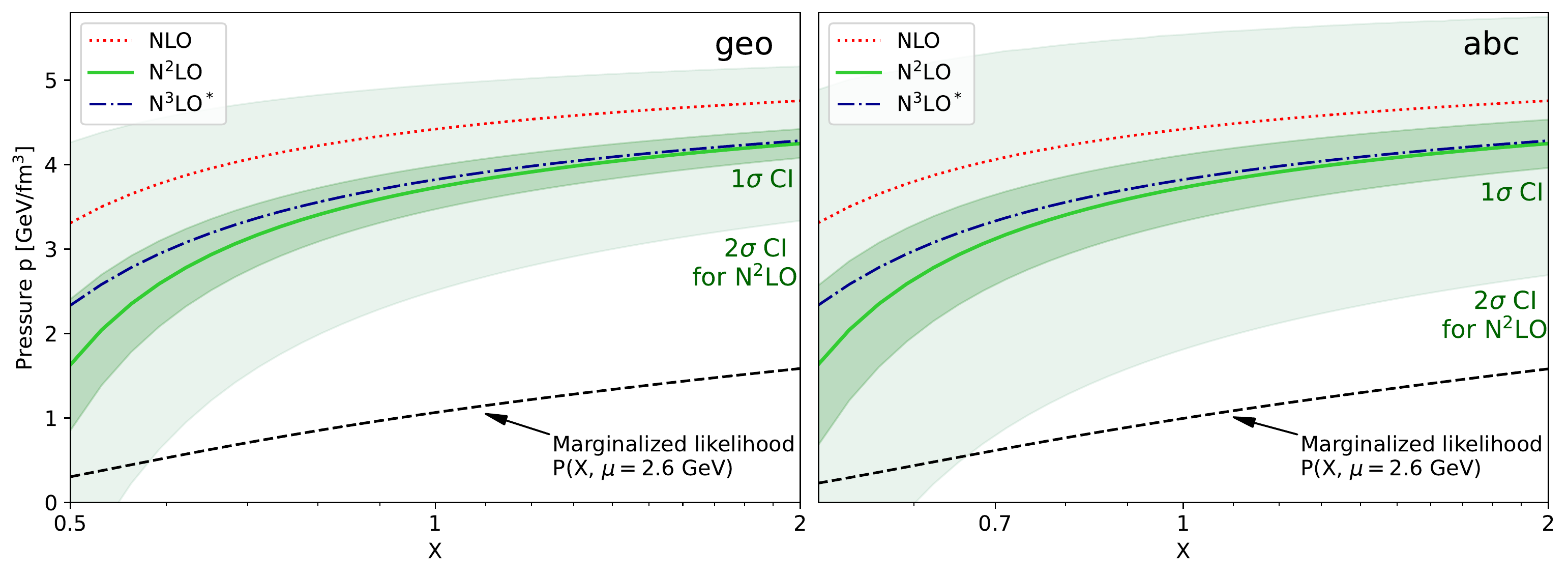}
    \caption{(Top) Order-by-order estimates of the pressure and its MHO uncertainty [\cref{eq:PMHO}] at $\mu = 2.6$~GeV with the renormalization scale parameter $X = 1$. The left panel assumes a geometrical model for the sequence of the perturbative corrections while in the right panel assumes an $abc$ model. (Bottom) 68\%- and 95\%-credible intervals ($1\sigma$ and $2\sigma$) predicted by the geometric and $abc$ models given the terms up to N$^2$LO as a function of $X$ at $\mu = 2.6$~GeV.}
    \label{fig:MHO_fixmedmu}
\end{figure*}

In the top panels of \cref{fig:MHO_fixmedmu} we show the posterior distributions for the pressure at fixed $\mu = 2.6$~GeV and $X = 1$ for the geometric (left) and $abc$ models (right). By construction, the LO distribution contains no useful information and is completely determined by priors. At NLO the distribution for the $abc$ model is asymmetric, in contrast to the geometric model. Because the NLO correction is negative, the $abc$ model infers an alternating series; hence, the distribution is skewed towards larger pressure values, i.e.,~it expects a positive N$^2$LO correction.  However, the actual N$^2$LO correction is again negative. This causes the $abc$-model posteriors to become symmetric.
In general both models predict similar posterior distributions with more input orders, as seen in the figure.
In the bottom panels of \cref{fig:MHO_fixmedmu} we show the 68\% and 95\% credible intervals (CIs) for the N$^2$LO posterior distributions.\footnote{Credible intervals are defined to contain the specified percentage of the probability distribution with the remaining probability split symmetry on either side of the CI.} We see that the 68\% CI fully incorporates the N$^3$LO${}^*$ corrections. The 95\% CI for the $abc$ model is noticeably wider than that of the geometric model. If the series were alternating as expected by the $abc$ model, then the posterior CIs would be narrower than the ones of the geometric model.

\subsection{Incorporating scale dependence}
\label{sec:MiHO_B}

\begin{figure*}
    \centering
    \includegraphics[width = 0.9 \textwidth]{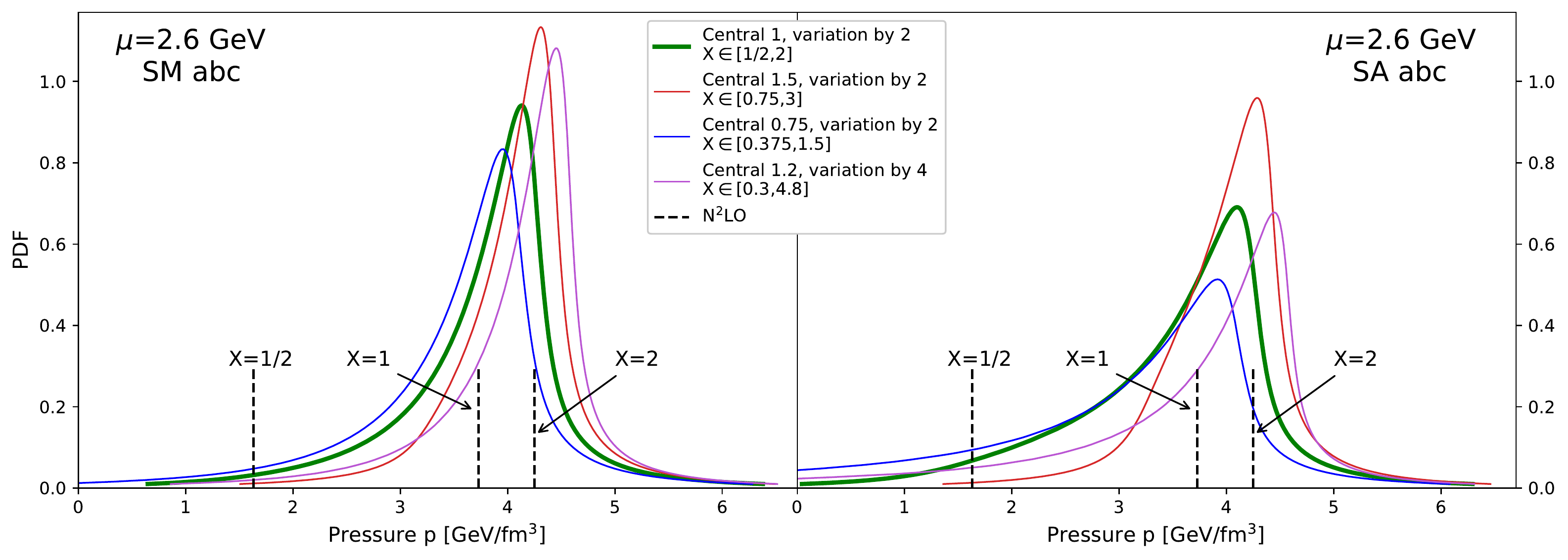}
    \caption{Estimation of the MHO uncertainty in pressure at $\mu=2.6$~GeV using (left) the scale marginalization prescription and (right) scale-averaging prescription, both using the $abc$ model for different ranges of the renormalization-scale parameter $X$. The black dashed lines correspond the N$^2$LO values of the pressure for different $X$. The green line corresponds to the central value of $X=1$ varied by a factor of 2. The red and blue lines demonstrate the effect of varying the central value ($X=0.75$ and $X=1.5$, respectively). The effect of the variation range is depicted by the violet line, where the renormalization scale is varied by a factor of 4 instead of 2. Only mild dependence on the choice of the range of $X$ is observed.}
    \label{fig:scales}
\end{figure*}

The truncated perturbative series depends on the unphysical renormalization scale $\bar{\Lambda}= \tfrac{2}{3}\mu X$, which is related to the characteristic hard scale of the process---the baryon number chemical potential $\mu$---via the dimensionless scaling factor $X$. 
Consequently,  the posterior probability of a Bayesian model is also sensitive to the choice of $X$. As there is no physical reason to prefer one scale choice over another, the scale dependence must be considered in estimating the observable's true value. 

The \textbf{scale-averaging} prescription advocated in \cite{Duhr:2021mfd} treats all scales on equal footing and adds them coherently:
\begin{equation}
    P_\text{sa}(\delta_{k + 1}| \bm{\delta}_k) \equiv \int dX P_0(X)  P(\delta_{k + 1}(X)| \bm{\delta}_k(X))\label{eq:sa}.
\end{equation}
Here the left-hand side should be understood as a scale-independent probability distribution for the N$^{(k+1)}$LO correction given terms up to N$^k$LO. The weight function 
\begin{equation}
P_0(X; X_0, F) \equiv \frac{1}{2 X F} \theta\left(\log F - \left|\log \frac{X}{X_0}\right|\right)
\end{equation}
with $F=2,X_0=1$ implements a log-uniform weight for the range $1/2<X<2$. Note that the relative weight of the distributions does not depend on the convergence of the series.

Alternatively, in the \textbf{scale-marginalization} prescription~\cite{Bonvini:2020xeo} the scale parameter $X$ is treated as a hidden model parameter. The scale-independent probability is obtained by marginalization over $X$
\begin{equation}
    P_\text{sm}(\delta_{k + 1}| \bm{\delta}_k) \equiv \int dX P(X|\bm{\delta}_k(X)) P(\delta_{k + 1}(X)| \bm{\delta}_k(X))\label{eq:sm},
\end{equation}
where the likelihood of $X$ given $\bm{\delta}_k$ is
\begin{equation}
   P(X|\bm{\delta}_k(X)) = \frac{ P_0(X)  P(\bm{\delta}_k(X))}{\int dX P_0(X)P(\bm{\delta}_k(X))}.
\end{equation}
Here $P(\bm{\delta}_k(X))$ is the marginalized likelihood (evidence) given by \cref{eq:evidence} and $P_0(X)$ is now interpreted as a prior for $X$.
The marginalized likelihood is the largest for values of $X$ for which the perturbative series is most convergent, i.e., favoring fastest apparent convergence \cite{Grunberg:1980ja,Stevenson:1981vj}. 
In the case of pressure, the convergence is the fastest for the largest values of $X$, so that the weighting with the marginalized likelihood slightly favours higher values of $X$, see \cref{fig:MHO_fixmedmu} for the marginalized likelihood as a function of $X$.
In contrast, for the scale-averaging prescription, the integral \cref{eq:sa} is dominated by the accumulation of probability distributions with slow $X$ dependence, i.e., it accords with the principle of minimal sensitivity \cite{Stevenson:1981vj}.

In \cref{fig:scales} we display the scale-marginalized distributions for the pressure given in \cref{eq:sm/sa}
at N$^2$LO with different choices for the central scale $X_0=\{1,1.5,0.75,1.2\}$ and the size of the marginalization window $F=\{2,2,2,4\}$ for a fixed $\mu = 2.6$~GeV. The position of the peak of distributions changes only by approximately 10\%. Notably, the perturbative series at low scale values does not affect the marginalized distribution. We also observe from this figure that the scale-marginalization prescription leads to distributions that are skewed to higher pressures than results of the naive scale-variation prescription, varying $X \in [1/2,2]$. Scale-averaged distributions are also biased towards larger $X$, because there the scale dependence is slowing down. However, in this case convergence properties are not taken into account and the resulting distributions are wider. 

\begin{figure}[h!]
    \centering
    \includegraphics[width=\columnwidth]{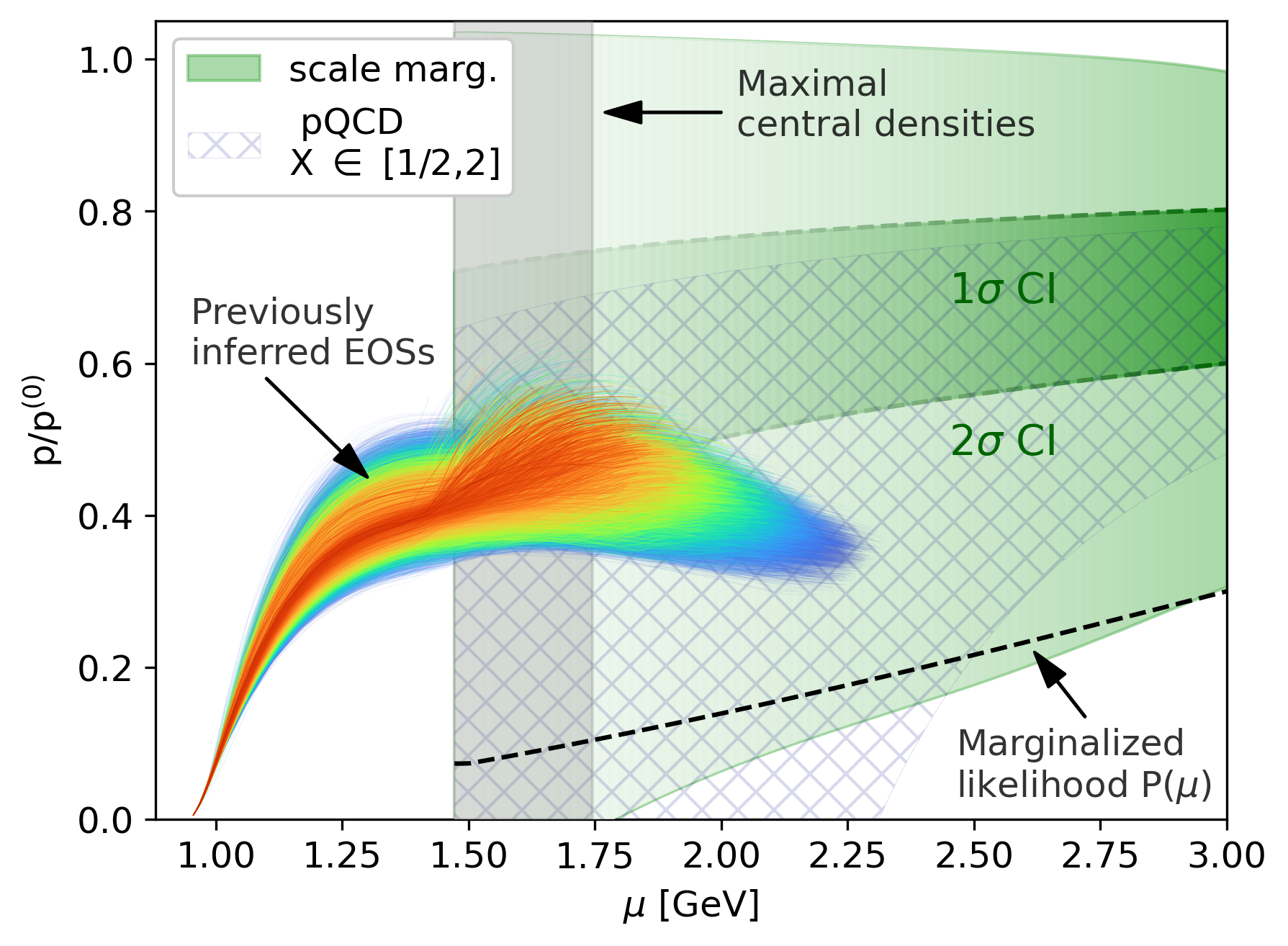}
    \caption{The pressure normalized by that of free Fermi gas of quarks as a function of chemical potential. The green bands correspond to the $\rm N^2LO$ pQCD calculations whose uncertainties are estimated by the $abc$ model using the scale-marginalization prescription for the renormalization scale $X$; the darker and the lighter bands represent 68\% and 95\%-credible intervals, respectively. The relative confidence in the $abc$ model is qualified by the marginalized likelihood $P(\mu)$ defined in \cref{eq:marg_likelihood} and illustrated by the black dashed line (and the fading of the green bands). The hatched purple band represents the standard error estimation of pQCD results obtained by renormalization scale variation by a factor of 2. Colored lines are the sample from the ensemble of NS EoSs used in \cite{Gorda:2022jvk} conditioned with astrophysical observations and QCD input for the scale-marginalization prescription for $X$ in the range $[1/2,2]$ and $\mu_{\rm QCD}$ in the range $[2.2, 3]$ GeV. The coloring of individual EoSs corresponds to the posterior likelihood. The higher likelihood is associated with darker shades of red.}
    \label{fig1}
\end{figure}

 In \cref{fig1} we can see how the scale-marginalized distribution of the pressure (green line in \cref{fig:scales}) changes with the chemical potential $\mu$.  The green bands show the $68\%$ and $95\%$ CIs. The 68\% CIs lies at the upper edge of the naive scale-variation estimate, shown in the figure as the hatched purple band. However the distributions have long power-law tails leading to 95\% CIs with larger bands than scale variation, especially for large $\mu$ values.

\subsection{\texorpdfstring{Comparison with Lattice-QCD calculations at finite $T$ and $\mu=0$}{Comparison with Lattice calculation at finite T and μ=0}}
\label{sec:high_T}

At high $T$ and $\mu = 0$, lattice computations of the EoS are available \cite{Borsanyi:2013bia}. This enables the comparison between the precise nonperturbative value of the pressure $p(T)$ and the Bayesian inference from the perturbative calculations. In the high-$T$ pQCD regime, due to the presence of thermally over-occupied, long-wavelength gluonic modes, the weak-coupling expansion for $p$ is structured as a power series in $g \equiv \sqrt{4 \pi \alpha_s}$, rather than in $\alpha_s$ itself. Hence, in this context the terms $p^{(k)}$ denote terms of order $O(g^k) = O(\alpha_s^{k/2})$, rather than $O(\alpha_s^k)$. Note however, that the hard contributions (from momenta of order $T$) are still structured as a power series in $\alpha_s \sim g^2$, since they arise from a finite number of individual bubble diagrams at the same naive order in the coupling. The resummed contributions on the other hand, arising from multiple Feynman diagrams featuring soft momenta $\sim \alpha^{1/2}_s T$ appear at all orders starting at $O(g^3)$. 
The $p^{(k)}$ are known up to $k = 5$, and can once again be extracted from \cite{Kurkela:2016was} (see \cref{sec:high_T_appendix}).

\begin{figure*}
    \centering
    \includegraphics[width = 0.9\textwidth]{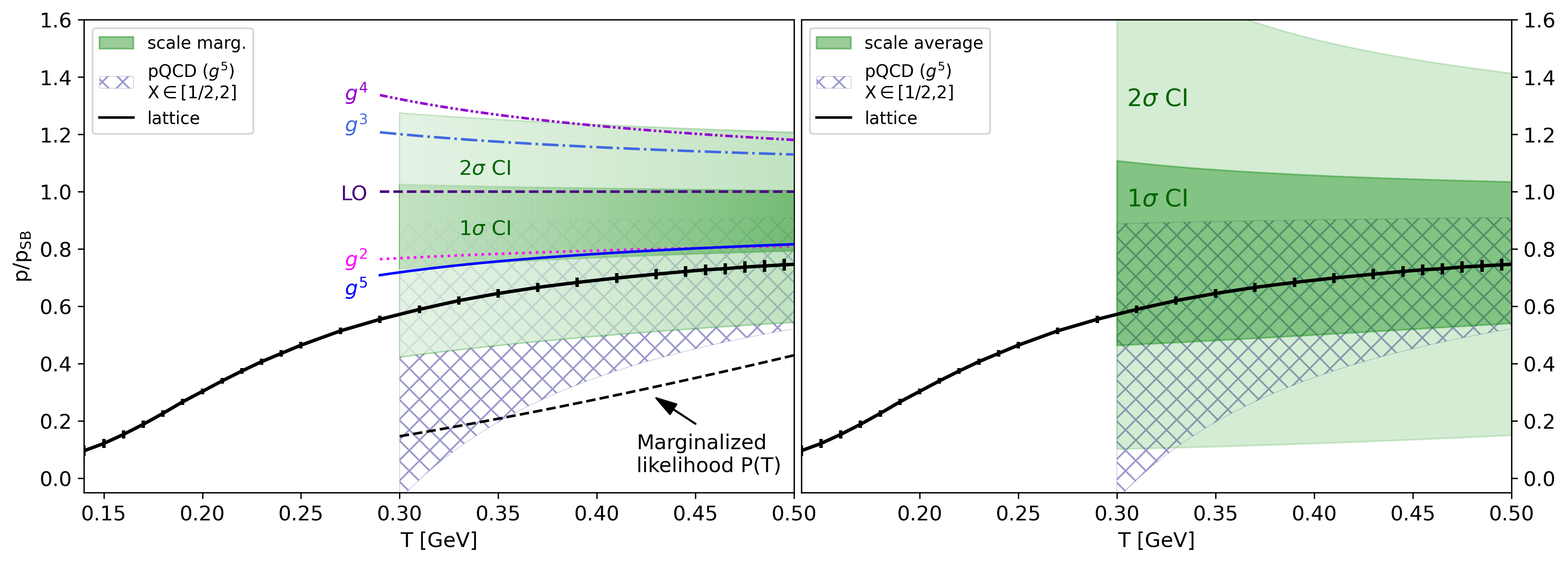}
    \caption{Applying the statistically interpretable error estimation to QCD thermodynamics at zero chemical potential. In this case the pressure can also be accessed non-perturbatively using lattice-field-theory methods \cite{Borsanyi:2013bia}. The green bands correspond to the 68\% and 95\%-CIs predicted by the $abc$ model using (left) the scale-marginalization  and (right) scale-averaging  prescription for $X$ in the range $[1/2,2]$. The HTL-resummed perturbative result from \cite{Kurkela:2016was} is shown with lines denoted by their order in the coupling constant $g$. The confidence 
    in the model 
    at various temperatures $T$ is characterized by the marginalized likelihood $P( T ) $ (displayed in arbitrary units).}
    \label{fig:finiteT}
\end{figure*}

As is well known \cite{Kajantie:2002wa} the perturbative expansion for the pressure at high-$T$ is not well behaved without additional non-trivial resummations \cite{Blaizot:2000fc,Laine:2006cp,Mogliacci:2013mca,Andersen:1999fw}. This can be readily seen from \cref{fig:finiteT}, where we show in different linestyles the different perturbative orders in the expansion. In particular, the result alternates above and below the free Stefan-Boltzmann (SB) value. In this case, due to the non-convergence of the series, the training of the statistical models is sensitive only to the $g^5$ term (i.e., if the model with given set of parameters can reproduce this term, it by construction can reproduce all the lower orders as well). Nevertheless, we show the CIs extracted with $abc$ model with scale-marginalization (left panel) and scale-averaging prescriptions (right panel). 
Scale-marginalization leads to narrower CIs and only the 95\% CI covers the non-perturbative results. Scale-averaging weights equally small $X$ values which go downwards at small temperatures. 

\section{Propagating pQCD to low densities}

We turn now to a discussion of how pQCD results at high densities constrain the EoS at densities that are relevant for NSs. We use the recently introduced prescription from \cite{Komoltsev:2021jzg} to determine the allowed region for the low-density EoS given pQCD predictions at some fixed large value of $\mu_H$. The choice of this reference scale $\mu_H$ is arbitrary as long as pQCD results are reliable. We will incorporate the uncertainty arising from the value of this scale into our predictions using a similar scale prescription as was done for the renormalization scale $\bar \Lambda$.

\subsection{ Constraints at low densities }
\label{sec:KoKu}
Knowledge of the EoS at high densities $n_H$ and baryon number chemical potentials $\mu = \mu_H$ where the pQCD calculation is convergent imposes robust constraints to lower densities $n_L$ reached in cores in NSs \cite{Komoltsev:2021jzg,Gorda:2022jvk}. The most conservative way of using the high-density information is to demand only that the EoS at lower densities can be connected to the EoS at $\mu_H$ using an interpolation that is mechanically stable, causal, and thermodynamically consistent.

Given the triplet of values at high densities 
\begin{align}
    \vec \beta_H(\mu_H, X)  \equiv  (p_H(\mu_H,X), n_H(\mu_H,X), \mu_H ),
\end{align}
we can impose a condition on the corresponding triplet at lower densities
$
    \vec \beta_L  = \bigl(p_L, n_L, \mu_L \bigr).
$
Specifically, if the difference in pressure $\Delta p \equiv p_H - p_L $ does not lie in the interval $[\Delta p_{\min} , \Delta p_{\max} ]$ where 
\begin{align}
    \Delta p_{\rm min} & \equiv \frac{c_{\rm s, lim}^2}{1 + c_{\rm s, lim}^2 }
    \left( \mu_H \left(\frac{\mu_H}{\mu_L}\right)^{1/c_{\rm s, lim}^2} - \mu_L \right) n_L,\\
     \Delta p_{\rm max} & \equiv \frac{c_{\rm s, lim}^2}{1 + c_{\rm s, lim}^2 }
    \left( \mu_H  - \mu_L \left(\frac{\mu_L}{\mu_H} \right)^{1/c_{\rm s, lim}^2} \right) n_H,
\end{align}
with $c^2_{\rm s, lim} = 1$,
then there is no mechanically stable, causal and thermodynamically consistent way to connect the EoS at low densities to the EoS at high densities. Therefore any EoS with $\Delta p \notin [\Delta p_\mathrm{min}, \Delta p_\mathrm{max}]$ is excluded by pQCD.

\begin{figure}
    \centering
    \includegraphics[width = \columnwidth]{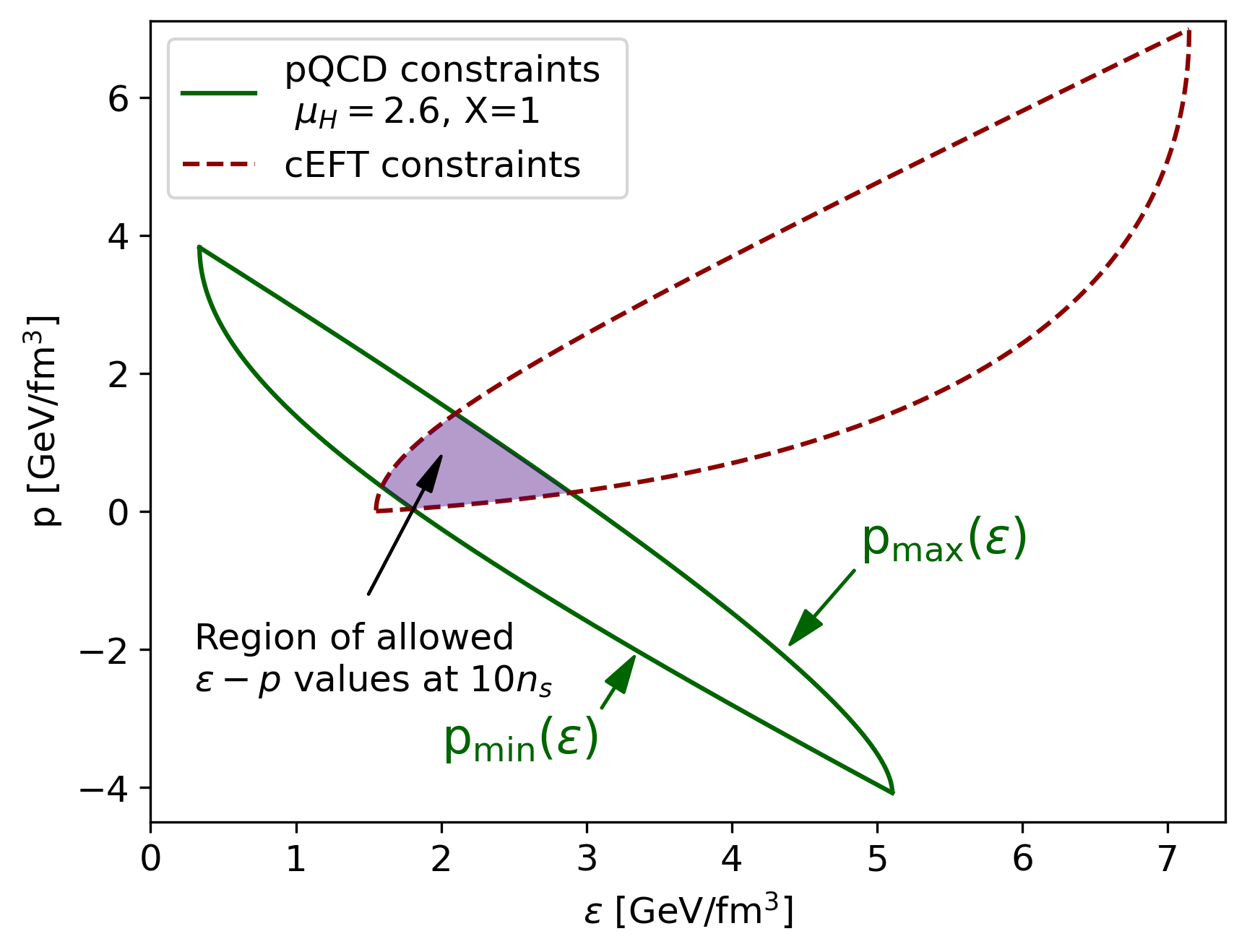}
    \caption{The constraints on the $\epsilon$--$p$ values arising from the requirement of the causal, mechanically stable and thermodynamically consistent EoS connected to the theoretical calculations. The red dashed region represents the allowed area if the EoS is connected only to the cEFT results averaged between stiff and soft from  \cite{Hebeler:2013nza}. The green line shows $p_{\min/\max}(\epsilon )$ [see \cref{eq:p_bounds}] indicating the region allowed by pQCD; the area bounded by the green line is given by \cref{eq:area}.  %
    }
    \label{fig:area_cet_qcd}
\end{figure}

These conditions restrict the area of allowed values the pressure $p_L$ and the energy density $\epsilon_L$ can take at the given low density $n_L$.  In particular, to be able to match to the triplet $\beta_H$, the energy density and pressure at the lower density $n_L$, must be bounded by the curves $p_{\min}(\epsilon) < p < p_{\max}(\epsilon)$
\begin{align}
    p_{\min}(\epsilon ) \equiv& p_H-\sqrt{\epsilon^2+2 \epsilon p_H-\mu_H^2 n_L^2+p_H^2} ,\nonumber \\
    p_{\max}(\epsilon ) \equiv& \frac{n_L \sqrt{\mu_H \left(-2 \epsilon n_H+\mu_H n_L^2+\mu_H n_H^2-2 n_H p_H\right)}}{n_H} \nonumber \\ 
    & -\epsilon+\frac{\mu_H n_L^2}{n_H},
\label{eq:p_bounds}
\end{align}
with $\epsilon$ being bounded by the intersections of the $p_{\min}$ and $p_{\max}$ curves denoted as $\epsilon_{\min}$ and $\epsilon_{\max}$
\begin{equation}
\label{eq:e_bounds}
\begin{split}
    \epsilon_{\min}\equiv&\mu_H n_L-p_H, \\
    \epsilon_{\max} \equiv&\frac{\mu_H \left(n_L^2+n_H^2\right)}{2 n_H}-p_H.
\end{split}
\end{equation}
See \cref{fig:area_cet_qcd} for illustration. In this figure, we also show the corresponding region of $p_L$ and $\epsilon_L$ values consistent with chiral effective field theory (cEFT) at low densities \cite{Hebeler:2013nza} (see \cref{sec:CEFT constraints} for the expressions for this corresponding boundary).
Note that negative values of pressure are allowed by pQCD since it is possible to draw casual and stable EoS from these point to high-density limit (they might be, however,  excluded for different reasons).

As the information from a fixed high-density triplet $\beta_H$ is propagated down to a lower density $n_L$, it gets spread over a wider and wider area in the $\epsilon$, $p$ plane. Specifically, the information becomes spread over the area bounded by the allowed curves in \cref{eq:p_bounds,eq:e_bounds}, namely 
\begin{align}
\label{eq:area}
   & A(\vec\beta_H, n_L)  \equiv \int^{\epsilon_{\max}}_{\epsilon_{\min}} d\epsilon \bigl[ p_{\rm max}(\epsilon)- p_{\rm min}(\epsilon) \bigr]\nonumber\\
    &=\frac{\mu _H^2 n_L}{12 n_H^2} \left(4 n_H^3-3 n_H^2 n_L-6 n_H^2 n_L \log \left(\frac{n_H}{n_L}\right)-n_L^3\right).
\end{align}
Now, the probability that a valid point $(\epsilon,  p)$ lies within the bounded region at $n_L$ is unity, so that 
\begin{align}
    1 = \int_{A(\vec\beta_H, n_L)} d\epsilon\, dp \frac{d^2 P(\epsilon, p | \vec\beta_H, n_L)}{d \epsilon \,d p},
\end{align}
where $d^2 P / (d \epsilon \, d p)$ is the differential probability of finding $p$ and $\epsilon$ in the element of area.  Taking the agnostic approach and not assigning higher likelihoods to less extreme EoSs, we may consider all allowed points equally likely, i.e.
\begin{align}
	\frac{d^2 P(\epsilon, p | \vec\beta_H, n_L)}{d \epsilon \, d p} = \textrm{const.} = 1/A(\vec\beta_H, n_L), 
\end{align}
where the constant is determined by the normalization condition of probability.

Therefore the conditional probability is
\begin{equation}
        \label{eq:KoKuWeight}
        P_{\rm \KoKu}(\epsilon_L, p_L |  n_L, \vec\beta_H) \equiv \frac{{\mathbf 1}_{[\Delta p_{\rm min}, \Delta p_{\rm max}]}(\Delta p) }{ A(\vec\beta_H, n_L) },
\end{equation}
where ${\mathbf 1}_{S}$ is the indicator function on the set $S$. Note that for fixed $\mu_H$ and $n_L$ the $X$ dependence of the area is mild,  
since the dependence $n_H (X) $ is mild and the more $X$-sensitive $p_H(X)$ does not appear in the expression for the area. 

\subsection{\texorpdfstring{Marginalization over $\mu_H$}{Marginalization over mu\_H}}
\label{sec:KoKu_B}

In \cite{Gorda:2022jvk} it was argued that one should use the perturbative information at the smallest possible density where MHO uncertainties are under control. However, the choice of $\mu_H$ is arbitrary as long as perturbative results are reliable. Hence, here we shall incorporate the $\mu_H$ dependence using the more sophisticated scale-marginalization prescription discussed above. Note that there are alternative ways to handle this $\mu_H$ dependence such as the scale-averaging prescription. %

In \cref{eq:KoKuWeight}, we introduced the conditional probability for the \KoKu{} construction in \cite{Komoltsev:2021jzg} weighted with the inverse area $A(\vec\beta_H, n_L)$. The dependence of this area on $\mu_H$ approximately scales as
\begin{equation*}
    A(\vec\beta_H, n_L) \sim \mu_H^5,
\end{equation*}
which follows from $n^{(0)}_H \sim \mu_H^3 $. 
Therefore, one can see that applying the QCD input at larger values of $\mu_H$ entails spreading the information over a larger allowed region of $\epsilon$--$p$ values. The area weight hence implies using smaller values of $\mu_H$ in order to increase the constraining power. The scale-marginalization prescription introduces the opposite behavior (see the marginalized likelihood $P(\mu)$ in \cref{fig1}), i.e., increasing the value of $\mu_H$ leads to more confidence in the perturbative calculation. This interplay allows a data-driven determination (based on ${\bm p}^{(k)}$) of the relevant range of chemical potential that is at the same time as constraining as possible while being perturbatively reliable.

For the scale-marginalization prescription over $\mu_H$ and $X$ we use the following integral weight 
\begin{align}
P_{\rm sm}(&\mu_H, X |\bm{p}^{(k)}, \bm{n}^{(k)}) = \nonumber \\
&\frac{ P_0(X) P_0(\mu_H) P(\bm{\delta}_k(X,\mu_H))}{\int dX d\mu_H P_0(X) P_0(\mu_H)P(\bm{\delta}_k(X,\mu_H))}\label{eq:KoKusm}.
\end{align}
Note that we use a common marginalized likelihood $P(\bm{\delta}_k(X,\mu_H))$ for both $X$ and $\mu_H$. This is justifiable because in the procedure dictated by \cite{Komoltsev:2021jzg} one compares predictions at the same low-density point derived from different series labeled by $\mu_H$ and $X$. Therefore \cref{eq:KoKusm} assigns weight according to the convergence properties, which depend both on $\mu_H$ and $X$. We can also define the marginalized likelihood over $X$ for $\mu_H$
\begin{align}
\label{eq:marg_likelihood}
    P(\mu_H) \equiv \int dX P_0(X)P(\bm{\delta}_k(X,\mu_H)),
\end{align}
which is shown in \cref{fig1} as the dashed black line.

\begin{figure*}
    \centering
    \includegraphics[width = 1.5\columnwidth]{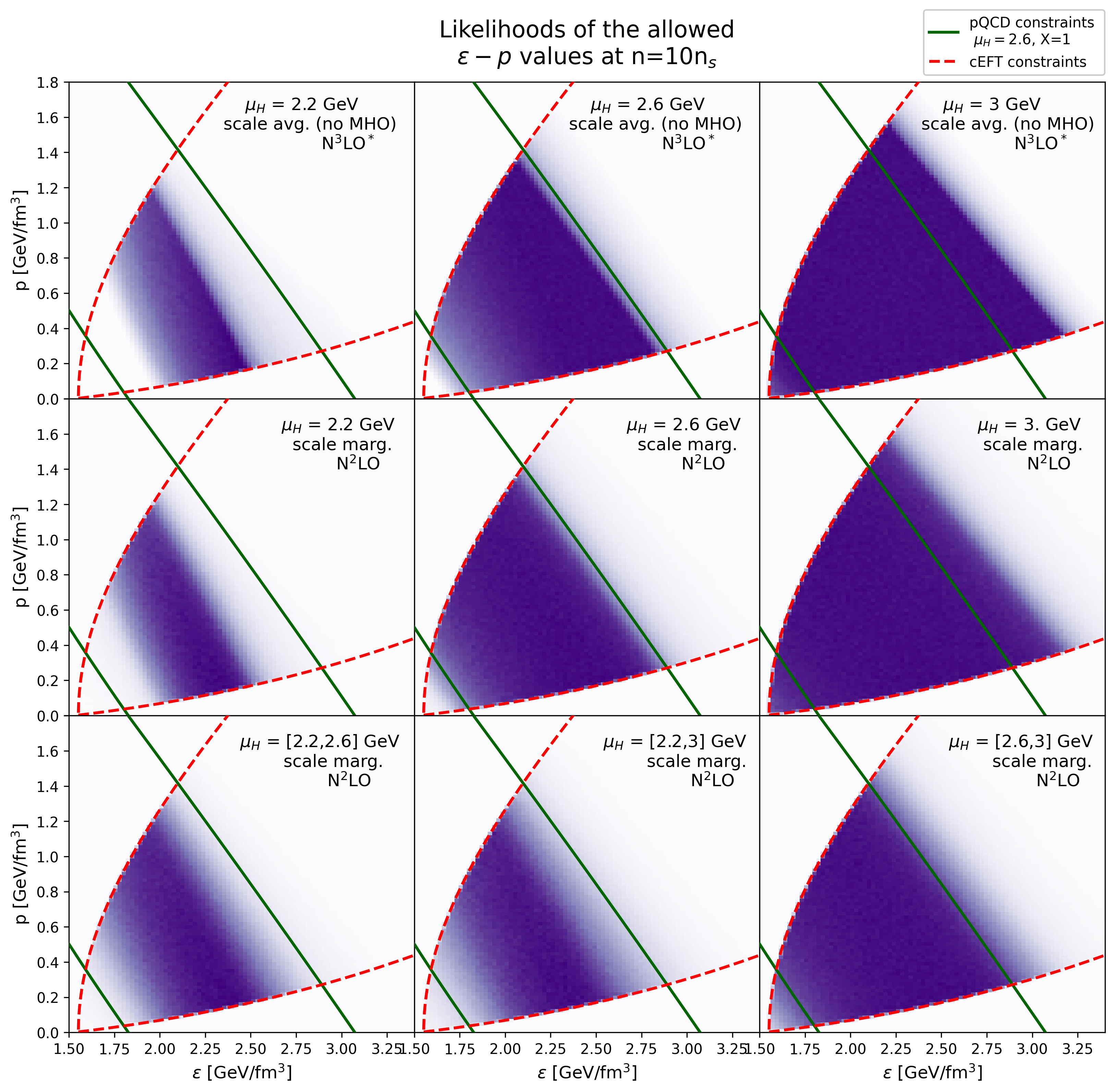}
    \caption{The panel of the likelihoods $P(\epsilon_L, p_L | n_L, \bm{p}^{(k)})$ within the allowed region of \cref{fig:area_cet_qcd} at fixed $n=10n_s$ using different prescriptions. The first row corresponds to the fixed $\mu_{H}$ = 2.2, 2.6 and 3 GeV and the log-uniform weight for $X$, and it assumes $p_H$ and $n_H$ are given by the N$^3$LO$^*$ pQCD values. The second row is obtained using the $abc$ model with scale-marginalization prescription for $X$ and fixed $\mu_{H}$. The third row introduces the simultaneous marginalization over $X$ and $\mu_{H}$ in the ranges $[2.2, 2.6]$, $[2.2, 3]$, $[2.6, 3]$ GeV. The solid green (the dashed red) lines are the same as in \cref{fig:area_cet_qcd} indicating the allowed region if EoS is connected to only pQCD (cEFT).}     
    \label{fig:panel}
\end{figure*}

The interplay between scale-marginalization and the area weight can be seen in \cref{fig:panel}. The panel of 9 likelihoods displays $P(\epsilon_L, p_L | n_L, \bm{p}^{(k)})$ calculated from \cref{eq:master_formula} at fixed $n_L=10n_s$ using different prescriptions for the $X$ and $\mu_H$ variations. The range $1/2\leq X \leq 2$ is fixed for all likelihoods. Note that in these figures we have assumed also that the EoS can be causally connected to the low-density nuclear EoS \cite{Hebeler:2013nza} in a mechanically stable, causal, and thermodynamically consistent way (i.e., we have restricted to the purple shaded region in \cref{fig:area_cet_qcd}); this is purely to highlight the relevant region of the QCD likelihood function. 

The first row reproduces the likelihood function used in \cite{Gorda:2022jvk}, where a scale-averaging procedure for $X$ was adopted, but $\mu_H$ was kept fixed. The MHO uncertainty was not explicitly considered and only the highest available perturbative order (N3LO*) was used, corresponding to replacing the MHO distribution in \cref{eq:jointPnp} with $\delta$-functions
\begin{align}
&P_{\rm MHO}( p_H, n_H | \bm{p}^{(k)}(\mu_H , X),\bm{n}^{(k)}(\mu_H , X))  \nonumber \\
&\quad \to \delta\Bigl(p_H - p_3^*(\mu_H, X) \Bigr)   \delta\Bigl(n_H - n_3^*(\mu_H, X)\Bigr) .
\end{align}
Going beyond \cite{Gorda:2022jvk} where only $\mu_H = 2.6$~GeV was used, we display in the first row results for three different chemical potentials: $\mu_H$ = 2.2, 2.6, and 3 GeV corresponding to $n_H \approx 23 n_s, 40 n_s$, and $63 n_s$.  The smooth boundary of the likelihood function arises from the scale variation of the $p_{\rm max}$ and $p_{\rm min}$ lines as a function of $X$. For lower $\mu_H$ values the allowed area is smaller as expected because $P_{\rm \KoKu}$ is more constraining; however, the boundaries are less sharp because the $X$-variation of the pressure is stronger.

The effect of the MHO and the scale-variation uncertainties for fixed $\mu_H$ is demonstrated in the second row. Here, the MHO uncertainty is estimated using the $abc$ model (using terms up to and including N$^2$LO) and we have marginalized over $X$. We see (consistent with speculation in \cite{Gorda:2022jvk}) that the MHO and scale-variation uncertainties have only a minor effect to the final likelihood function; the main effect is to somewhat further blur the $p_{\rm max}$ and $p_{\rm min}$ boundaries compared to the first row. 

Lastly, the third row constitutes a main result of this paper, in which the final QCD likelihood function using the master formula \cref{eq:master_formula} is displayed. Compared to the second row, it includes the simultaneous marginalization over $X$ and $\mu_{H}$ from \cref{eq:KoKusm} with different $\mu_{H}$ ranges. We observe that the interplay between the area weight and the marginalized likelihood makes the more aggressive range with lower values of $\mu_H \in [2.2, 2.6]$~GeV (lower left panel) rather consistent with the more agnostic range $\mu_H \in  [2.2, 3]$~GeV (lower center panel). In particular it demonstrates the insensitivity to the upper limit of the $\mu_H$ range.  The most conservative range, $\mu_H \in [2.6, 3]$~GeV (right lower panel), shows a somewhat larger allowed area in the $\epsilon$--$p$ plane; we note that this most conservative choice produces a likelihood rather close to the likelihood function used in \cite{Gorda:2022jvk}, given by the $\mu_H = 2.6$~GeV panel in the top row.

\section{Discussion and conclusions}

In this paper, we have systematically discussed the effect of uncertainties of the perturbative-QCD input on the theoretically allowed $\epsilon$--$p$ region at neutron-star densities ($n_L = 10n_s$) arising from the missing higher-order terms, the renormalization-scale variation, and the choice of reference density. To this end we employed Bayesian-inference techniques developed in high-energy physics~\cite{Cacciari:2011ze,Bonvini:2020xeo,Duhr:2021mfd}. The main results of this inference are displayed in the bottom row of \cref{fig:panel}, from which we conclude that pQCD constraints are robust with respect to different sources of uncertainties.

\begin{figure}
    \centering
    \includegraphics[width = \columnwidth]{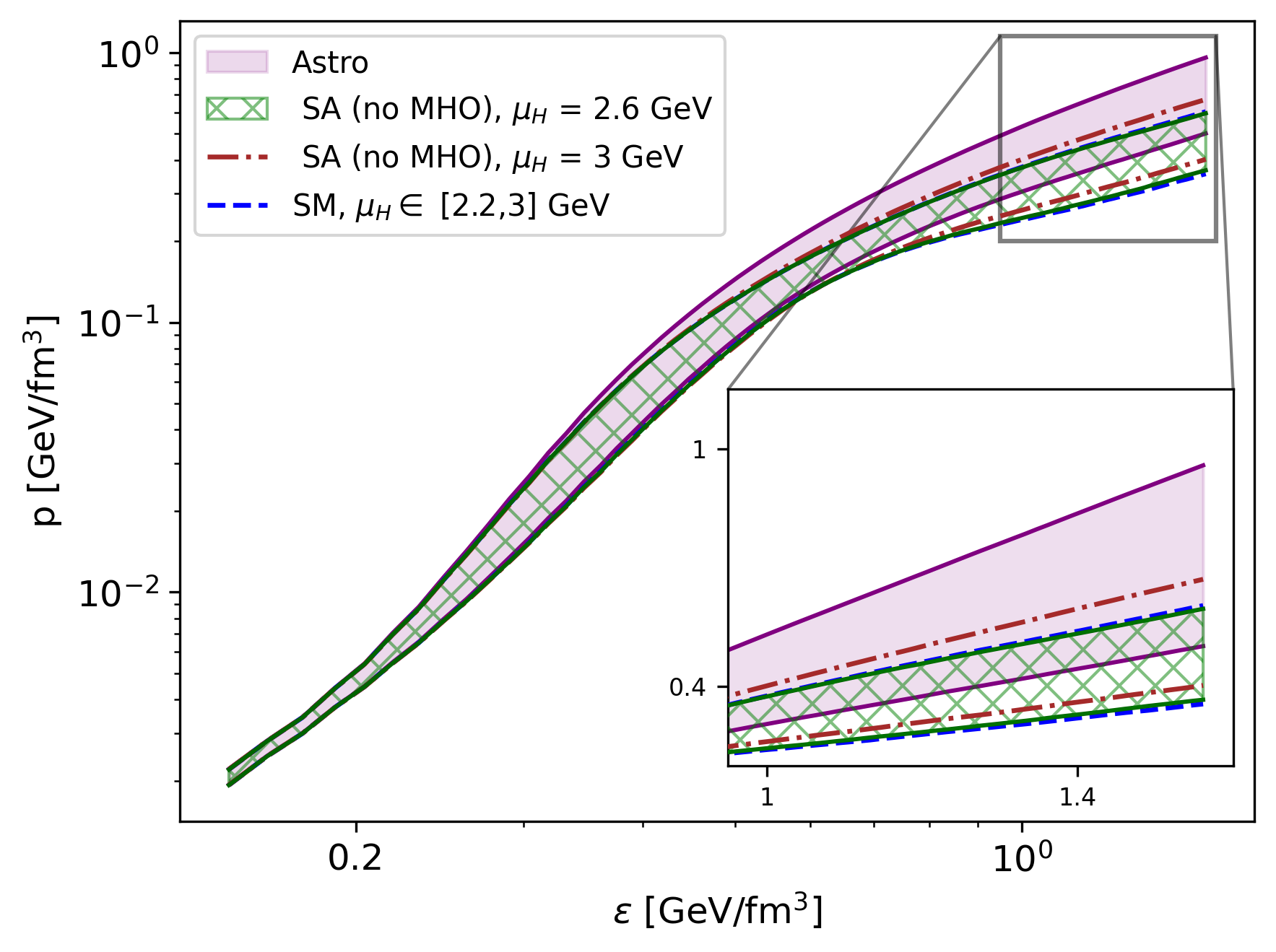}
    \caption{The impact of the QCD input using different prescriptions on the EoS inference. The bands correspond to 68\%-credible intervals. The ensemble conditioned only with astrophysical observations and cEFT is shown in pink. The green and the red bands correspond to the QCD input obtained using the log-uniform weight for $X$ in the range [1/2,2] with QCD imposed at fixed $\mu_{H}=$ 2.6 and 3 GeV, respectively. The blue dashed band represents QCD input obtained using the prediction of the $abc$ model with scale-marginalization prescription for $X$ and $\mu_{H}$ in range [2.2, 3] GeV. Overall, the dependence on the ranges and the prescriptions is mild.}
    \label{fig:ep}
\end{figure}

It remains a pertinent question how much the improved pQCD uncertainty estimation affects the global equation-of-state (EoS) inference of neutron-star matter.
To provide the first insights on this topic we interface the QCD likelihood function obtained here with the EoS inference setup of \cite{Gorda:2022jvk}. 
All the details of the EoS inference strictly follow those of \cite{Gorda:2022jvk} with the exception of the QCD likelihood function used. 
In \cref{fig:ep} we show the effect of the pQCD input on the posterior EoS region that has already been conditioned using a set of astrophysical and cEFT constraints, shown in pink. The green hatched band shows the effect
for a fixed value of $\mu_H$ = 2.6 GeV with log-uniform weights for $X \in [1/2, 2]$, which was denoted as ``Pulsars+$\tilde\Lambda$+QCD'' in \cite{Gorda:2022jvk}. The red dot-dashed band is the corresponding result when QCD is imposed at a larger chemical potential of $\mu_H$ = 3 GeV, which is the most conservative choice discussed here. The blue dashed line corresponds to the case of scale marginalizing $\mu_H$ in the range from 2.2 to 3 GeV and $X$ in the range from  1/2 to 2 obtained using the $abc$ model, which is the most comprehensive and agnostic uncertainty estimation we consider. 
We see that in all three of these cases, the pQCD input softens the EoS at high densities. Marginalizing over $\mu_H$ only slightly widens the band at the high energy densities. Even though we allow $\mu_H$ as high as 3.0 GeV (or $n\approx 64 n_s$) the band is only mildly affected.

In \cref{fig1} in the introduction we have already shown EoSs in the form of pressure as a function of chemical potential, where the likelihood has been computed using the prescription of the blue dashed lines in \cref{fig:ep} (with the same astrophysical constraints as above). We see from \cref{fig1} that the QCD EoS constrained by astrophysical observations at low densities can be smoothly connected to our posterior distribution of the pQCD pressure at higher densities. Remarkably, even at the maximal densities reached within stable NSs, the EoS is in rough agreement with the 68\% CIs from the posterior distribution.

We conclude that the effect of the pQCD input on the NS-EoS inference is insensitive to the different prescriptions and choices used to estimate the uncertainties in the pQCD calculation. 
This increases the confidence in their use within neutron-star EoS inference.

\acknowledgements
We thank Alexander Huss, Tore Kleppe, Jan-Terje Kval\o y, Risto Paatelainen, Saga S\"appi, Achim Schwenk, and Aleksi Vuorinen for helpful discussions and comments at various stages of this project. This work is supported in part by the Deutsche Forschungsgemeinschaft (DFG, German Research Foundation)--project number 279384907--SFB 1245 (T.G.) and project number 496831614 (A.M.) and by the State of Hesse within the Research Cluster ELEMENTS (Project ID 500/10.006) (T.G.).

\appendix

\section{\texorpdfstring{pQCD terms at high $T$ and $\mu = 0$}{pQCD terms at high T and mu = 0}}
\label{sec:high_T_appendix}

In this section, we provide the perturbative terms used in \cref{sec:high_T}. We remind the reader that in this section, the terms of order $k$ scale as $O(g^k) = O(\alpha^{k/2}_s)$. The perturbative terms are \cite{Kurkela:2016was}
\begin{align}
    p^{(0)} =& \;\frac{19 \pi^2 T^4}{36}, \\
    p^{(1)} = & 0 \\
    p^{(2)} =& -\frac{3}{8} g(\bar\Lambda)^2 T^4.
\end{align}
Beginning with $p^{(3)}$, soft contributions appear in the form of terms arising from dimensional reduction (DR) as well as hard-thermal-loop (HTL) resummation:
\begin{align}
    p^{(3)} =& p^{(3)}_\mathrm{soft, DR}, \\
    p^{(4)} =& p^{(4)}_\mathrm{hard} + p^{(4)}_\mathrm{soft, DR} + p^{(4)}_\mathrm{soft, HTL}, \\
    p^{(5)} =&  p^{(5)}_\mathrm{soft, DR}. 
\end{align}
These terms take the form 
\begin{align}
    p^{(3)}_\mathrm{soft,DR} =\,\,\,& \frac{d_A m_E^3 T}{12\pi}\\
    p^{(4)}_\mathrm{hard} =& \frac{g(\bar\Lambda)^4}{(4\pi)^2} \left[ \frac{779}{48}-378 \zeta '(-1)-810 \zeta '(-3)\right. \nonumber \\
    &\left. +\frac{117}{4} \log\! \left( \frac{\bar\Lambda}{4 \pi T} \right) +\frac{9 \gamma_E }{2}-\frac{403 \log (2)}{60} \right] T^4 \\
    p^{(4)}_\mathrm{soft,DR} =& -\frac{g(\bar\Lambda)^2 m_E^2 T^2}{(4 \pi )^2}d_A C_A\left[\log \left(\frac{\bar\Lambda}{2 m_E} \right)+\frac{3}{4}\right] \\
    p^{(4)}_\mathrm{soft,HTL} =& \frac{d_A m_E^4}{256 \pi^2} f_\mathrm{HTL} ( T / m_E ) \\
    p^{(5)}_\mathrm{soft,DR}
    =
    &-\frac{C_A^2 d_A  g(\bar\Lambda)^4 m_E T^3}{(4 \pi )^3} \left[ \frac{\pi ^2}{6}+\frac{89}{24}-\frac{11}{6} \log (2)\right] \nonumber \\
    &+ \frac{d_A g(\bar\Lambda)^5 T^4}{128 \pi^3} \sqrt{\frac{3}{2}} \biggl[ 27 \log \left(\frac{\bar\Lambda}{4 \pi  T}\right)+\frac{13}{2} \nonumber\\
    & +27 \gamma_E -12 \log (2) \biggr] 
\end{align}
Here, $m_E^2 = \frac{3 g(\bar\Lambda)^2 T^2}{2}$ is the screening mass scale, and $f_\mathrm{HTL}$ is a numerical function discussed in \cite{Kurkela:2016was}. When we use these results, we use the same two-loop running coupling as used above in previous sections, and we take as the central value of the renormalization scale by $\bar\Lambda = 0.723 \times 4\pi T$, following \cite{Kajantie:2002wa}, in place of the $2 \mu_q$ used at $T = 0$.

\section{Propagating cEFT to higher densities}
\label{sec:CEFT constraints}

Similar to how the pQCD calculations at high $\mu_H$ restrict the range of intermediate energy densities and pressures at a fixed density $n_L$ using only the conditions of stability, causality and thermodynamic consistency, cEFT calculations at low densities also restrict the range of $p_L$ and $\epsilon_L$ values at higher densities. Given a triplet of cEFT values
\begin{equation}
    \vec{\beta}_{\ceft} \equiv (p_{\ceft}, n_{\ceft}, \mu_{\ceft}),
\end{equation}
the energy density and pressure at a fixed $n_L > n_{\ceft}$ must be bounded by the curves
 $p^{\ceft}_{\min}(\epsilon) < p < p^{\ceft}_{\max}(\epsilon)$
\begin{align}
    p^{\ceft}_{\min}(\epsilon) \equiv& 
    - \frac{n_L \sqrt{\mu_{\ceft} \left(-2 \epsilon n_{\ceft}+\mu_{\ceft} n_L^2+\mu_{\ceft} n_{\ceft}^2-2 n_{\ceft} p_{\ceft}\right)}}{n_{\ceft}} \nonumber \\ 
    & -\epsilon+\frac{\mu_{\ceft} n_L^2}{n_{\ceft}}, \\
    p^{\ceft}_{\max}(\epsilon ) \equiv& p_{\ceft}+\sqrt{\epsilon^2+2 \epsilon p_{\ceft}-\mu_{\ceft}^2 n_L^2+p_{\ceft}^2} ,
\label{eq:pCEFT_bounds}
\end{align}
with $\epsilon$ being bounded by the intersections of the $p^{\ceft}_{\min}$ and $p^{\ceft}_{\max}$ curves denoted as $\epsilon^{\ceft}_{\min}$ and $\epsilon^{\ceft}_{\max}$
\begin{equation}
\label{eq:eCEFT_bounds}
\begin{split}
    \epsilon^{\ceft}_{\min}\equiv&\mu_{\ceft} n_L-p_{\ceft}, \\
    \epsilon^{\ceft}_{\max} \equiv&\frac{\mu_{\ceft} \left(n_L^2+n_{\ceft}^2\right)}{2 n_{\ceft}}-p_{\ceft}.
\end{split}
\end{equation}
The area bounded by these curves is given by
\begin{align}
\label{eq:CEFT_area}
   & A_{\ceft}(\vec\beta_{\ceft}, n_L)  \equiv \int^{\epsilon^{\ceft}_{\max}}_{\epsilon^{\ceft}_{\min}} d\epsilon \bigl[ p^{\ceft}_{\rm max}(\epsilon)- p^{\ceft}_{\rm min}(\epsilon) \bigr]\nonumber\\
    &=-\frac{\mu _{\ceft}^2 n_L}{12 n_{\ceft}^2} \left(4 n_{\ceft}^3-3 n_{\ceft}^2 n_L-6 n_{\ceft}^2 n_L \log \left(\frac{n_{\ceft}}{n_L}\right)-n_L^3\right).
\end{align}

\bibliography{main}
\end{document}